\documentclass[a4paper,aps,twocolumn,superscriptaddress]{revtex4}%
\usepackage{amsfonts}
\usepackage{amsmath}
\usepackage{amssymb}
\usepackage{graphicx}%
\usepackage[dvipsnames]{xcolor}
\newcommand{\beginsupplement}{%
        \setcounter{table}{0}
        \renewcommand{\thetable}{S\arabic{table}}%
        \setcounter{figure}{0}
        \renewcommand{\thefigure}{S\arabic{figure}}%
}

\begin{document}

\title{Lifshitz transition and frustration of magnetic moments in infinite-layer NdNiO$_2$ upon hole-doping}

\author{I. Leonov}
\affiliation{M.N. Miheev Institute of Metal Physics, Russian Academy of Sciences, 620108 Yekaterinburg, Russia}
\affiliation{Ural Federal University, 620002 Yekaterinburg, Russia}

\author{S. L. Skornyakov}
\affiliation{M.N. Miheev Institute of Metal Physics, Russian Academy of Sciences, 620108 Yekaterinburg, Russia}
\affiliation{Ural Federal University, 620002 Yekaterinburg, Russia}

\author{S. Y. Savrasov}
\affiliation{University of California, Davis}

\begin{abstract}
Motivated by the recent discovery of superconductivity in the infinite-layer (Sr,Nd)NiO$_2$ films with 
Sr content $x \simeq0.2$ [Li et al., Nature (London) \textbf{572}, 624 (2019)], we examine the effects 
of electron correlations and Sr-doping on the electronic structure, Fermi surface topology, and magnetic 
correlations in (Nd,Sr)NiO$_2$ using a combination of dynamical mean-field theory of correlated 
electrons and band-structure methods. Our results reveal a remarkable orbital selective renormalization 
of the Ni $3d$ bands, with $m$*/$m\sim 3$ and 1.3 for the $d_{x^2-y^2}$ and $d_{3z^2-r^2}$ orbitals, 
respectively, that suggests orbital-dependent localization of the Ni $3d$ states.
We find that upon hole doping (Nd,Sr)NiO$_2$ undergoes a Lifshitz transition of the Fermi surface which 
is accompanied by a change of magnetic correlations from the three-dimensional (3D) N\'eel $G$-type 
(111) to the quasi-2D $C$-type (110). 
We show that magnetic interactions in (Nd,Sr)NiO$_2$ demonstrate an unanticipated frustration, which 
suppresses magnetic order, implying the importance of in-plane spin fluctuations to explain its 
superconductivity. 
Our results suggest that frustration is maximal for Sr-doping $x \simeq 0.1$--0.2, which is in agreement with an experimentally observed doping value Sr $x \simeq 0.2$ of superconducting (Nd,Sr)NiO$_2$. 
%

\end{abstract}

\maketitle


The recent discovery of superconductivity in the infinite-layer Sr-doped NdNiO$_2$ films 
(Nd$_{0.8}$Sr$_{0.2}$NiO$_2$) with the critical temperature up to $T_c \sim 15$ K has attracted a lot 
of attention from researchers around the world \cite{Nature.572.624}. NdNiO$_2$ has a similar planar 
crystal structure to that of the parent ``infinite layer'' superconductor CaCuO$_2$, which exhibits 
superconductivity below $T_c \simeq 110$ K upon hole doping \cite{Nature.356.775,NatPhys.13.1201,PhysRevLett.77.4430}. 
As Ni is isoelectronic to copper in NdNiO$_2$, it has a nominal $d^9$ configuration. Based on this, 
it was expected that analogous to cuprates the low energy physics of Sr-doped NdNiO$_2$ is dominated 
by electrons in the planar Ni $x^2-y^2$ states.
However, unlike cuprates, in the infinite-layer nickelate the Ni $x^2-y^2$ states are found to 
experience strong hybridization with the Nd $5d$ orbitals (primarily the $3z^2-r^2$ and $xy$ 
orbitals), yielding a non-cuprate-like Fermi surface \cite{Anisimov+Pickett,NatMater2020}.
While the electronic structure of NdNiO$_2$ has recently been widely studied using various band structure methods \cite{DFT_appl,PhysRevX.10.011024,Geisler_arxiv_2020}, model techniques \cite{Models_appl,Sakakibara_arxiv_2019}, and DFT+dynamical mean-field theory (DFT+DMFT) \cite{dmft,dft+dmft} methods \cite{PhysRevB.101.041104, Ryee_arxiv_2019, Gu_arxiv_2019, Held_group, Lechermann_arxiv_2019, Karp_arxiv_2020}, the properties of Sr-doped NdNiO$_2$ are still poorly understood. For NdNiO$_2$, DFT+DMFT calculations reveal significant correlation effects within the Ni $3d$ orbitals, which are complicated by large hybridization with the Nd $5d$ states \cite{Lechermann_arxiv_2019,PhysRevB.101.041104}.
Moreover, based on the experiments two features that are central to copper oxides -- the Zhang–Rice singlet and large planar spin fluctuations -- were claimed to be absent (or diminished) in (Nd,Sr)NiO$_2$  \cite{Nature.572.624,NatMater2020}.
%

Here we explore the effects of electronic correlations and Sr-doping on the electronic structure of (Nd,Sr)NiO$_2$ using a fully self-consistent in charge density DFT+DMFT method \cite{dmft,dft+dmft} implemented with plane-wave pseudopotentials \cite{pseudo,KCuF3}. 
DFT+DMFT has been proved to be among the most advanced theoretical methods for studying the electronic 
properties of strongly correlated materials, such as correlated 
transition metal oxides, heavy-fermions, Fe-based superconductors, e.g., to study the phenomena of a 
Mott transition, collapse of local moments, large orbital-dependent renormalizations, 
etc.~\cite{dftdmft_application} We use this advanced computational method to study the Fermi surface topology and magnetic correlations, as well as their impact on magnetism of (Nd,Sr)NiO$_2$ upon Sr-doping.

We adopt the experimental lattice parameters measured for the Nd$_{0.8}$Sr$_{0.2}$NiO$_2$ film grown on the SrTiO$_3$ substrate (space group $P4/mmm$, lattice parameters $a=3.91$ \AA\ and $c=3.37$ \AA) \cite{Nature.572.624}.
Following the literature, to avoid the numerical instabilities arising from the Nd $4f$ electrons we focus on La$^{3+}$ instead of Nd$^{3+}$ ($4f^3$) ion \cite{PhysRevX.10.011024,Sakakibara_arxiv_2019,Ryee_arxiv_2019}. (Hereafter, we assume La by saying Nd in our calculations).
%
%
To explore the effect of Sr-doping on the electronic structure of (Nd,Sr)NiO$_2$ we employ a rigid-band shift of the Fermi level within DFT.
In our DFT+DMFT calculations we explicitly include the Ni $3d$, Nd $5d$, and O $2p$ valence states, by constructing a basis set of atomic-centered Wannier functions within the energy window spanned by these bands \cite{Wannier}. 
This allows us to take into account a charge transfer between the partially occupied Ni $3d$, Nd $5d$, and O $2p$ states, accompanied by the strong on-site Coulomb correlations of the Ni $3d$ electrons.
We use the continuous-time hybridization expansion (segment) quantum Monte Carlo algorithm in order to solve the realistic many-body problem \cite{CT-QMC}. We take the average Hubbard $U = 6$ eV and Hund's exchange $J = 0.95$ eV as previously employed for rare-earth nickelates $R$NiO$_3$ \cite{Park2014}. We use the fully localized double-counting correction, evaluated from the self-consistently determined local occupations, to account for the electronic interactions already described by DFT.

\begin{figure}[h] 
\includegraphics[width=0.4\textwidth]{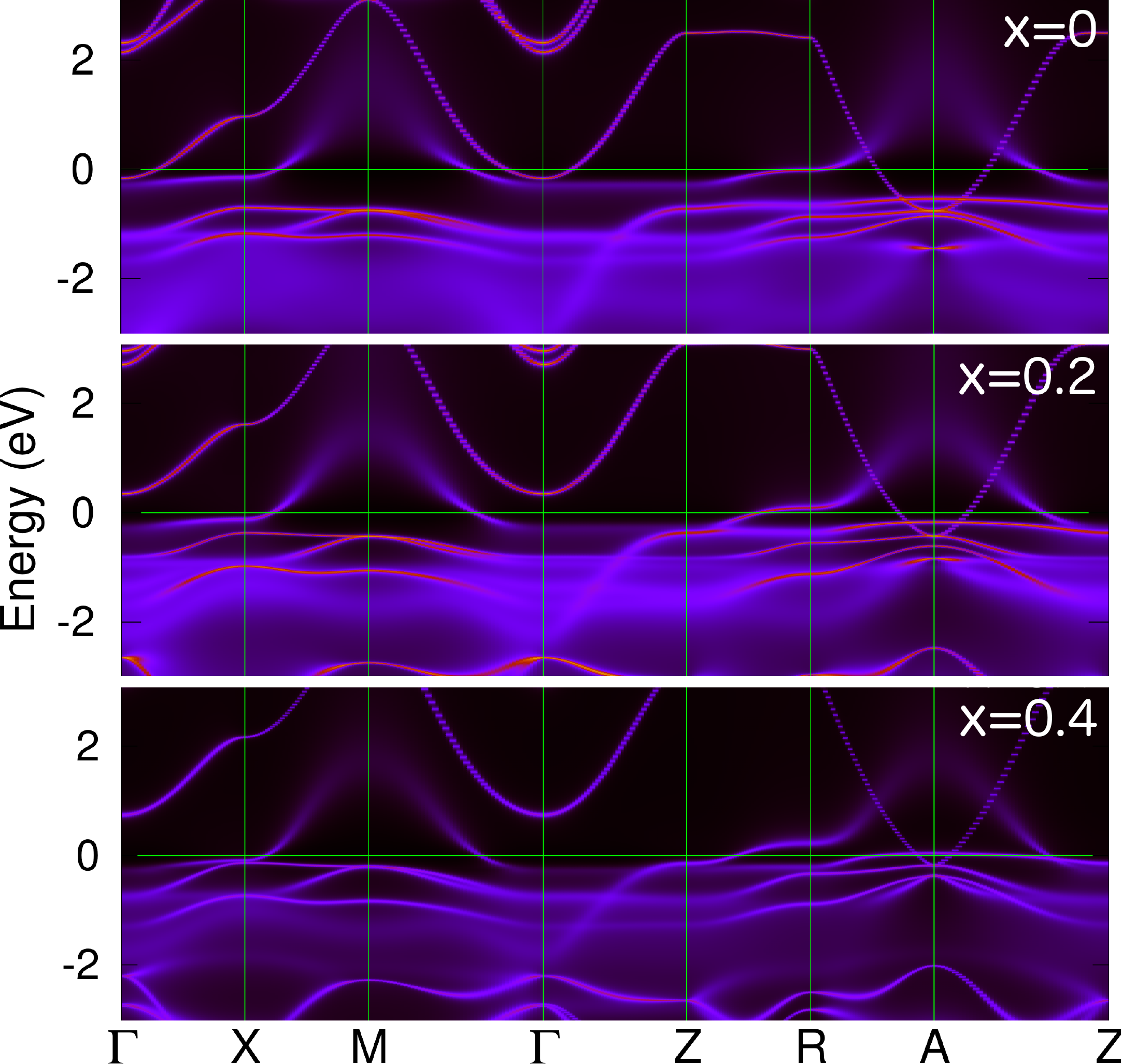}
\caption{{\bf k}-resolved spectral function of Sr-doped NdNiO$_2$ for Sr $x=0$ (top), $x=0.2$ (middle) and $x=0.4$ (bottom) as obtained by DFT+DMFT for the paramagnetic state at $T=290$ K.}
\label{fig:band_structure}
\end{figure}

In Fig.~\ref{fig:band_structure} we display our results for the {\bf k}-resolved spectra of paramagnetic (PM) (Nd,Sr)NiO$_2$ obtained by DFT+DMFT as a function of Sr doping $x$. Overall, our results agree well with those published previously \cite{Lechermann_arxiv_2019,Ryee_arxiv_2019,Sakakibara_arxiv_2019}. For Sr $x=0$ we observe a band formed by the strongly mixed Ni and Nd $3z^2-r^2$ states crossing the Fermi level near the $\Gamma$ point.
Upon Sr $x$ (hole) doping these states are seen to shift above the Fermi level, resulting in a change of the electronic structure of (Nd,Sr)NiO$_2$.

\begin{figure}[h]
\includegraphics[width=0.4\textwidth]{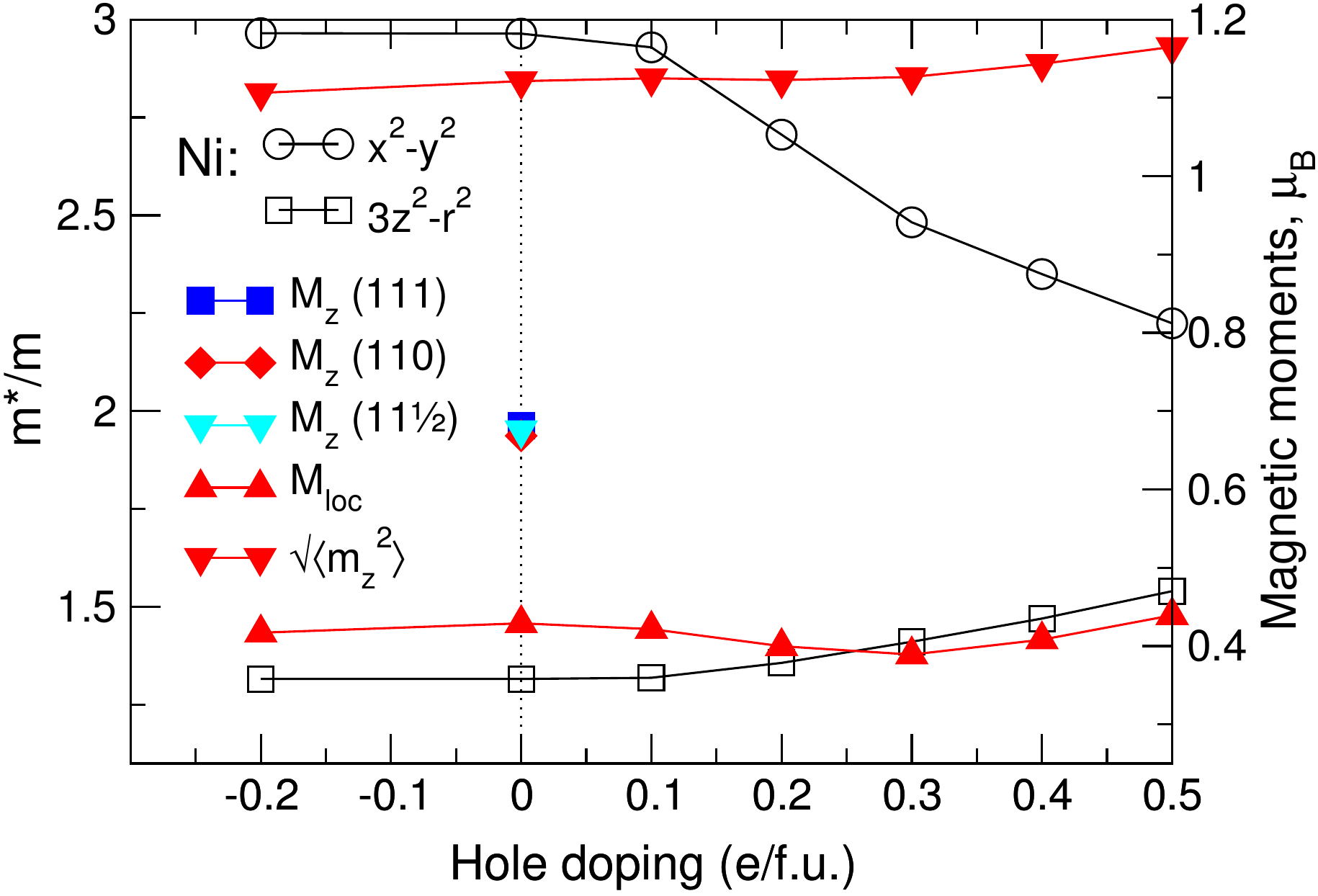}
\caption{Orbitally resolved quasiparticle mass enhancement $m$*/$m$ together with the instantaneous $\sqrt{\langle \hat{m}_z^2\rangle}$ and fluctuating magnetic moments $M_\mathrm{loc}=[T \int_0^{1/T} \langle \hat{m}_z( \tau) \hat{m}_z(0) \rangle]^{1/2}$ of Sr-doped NdNiO$_2$ calculated by DFT+DMFT for the paramagnetic state, at $T=290$ K. $M_z$: DFT+DMFT results for magnetization per Ni site for the N\'eel $(111)$, $C$-type $(110)$, and $(11\frac{1}{2})$ AFM states at $T=290$ K.}
\label{fig:correlations}
\end{figure}

\begin{figure}[h]
\includegraphics[width=0.4\textwidth]{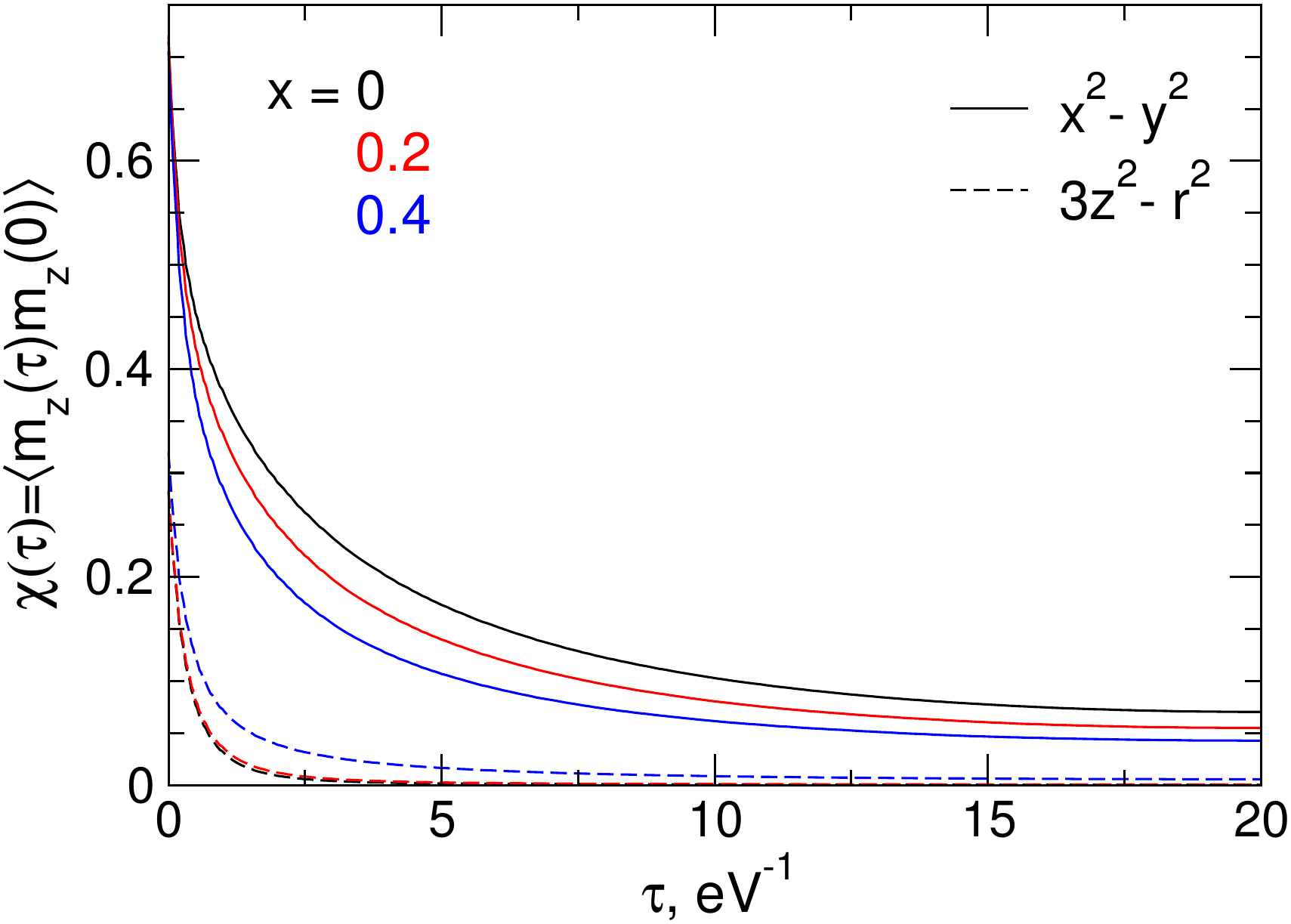}
\caption{Orbitally resolved local spin correlation functions $\chi(\tau) = \langle \hat{m}_z(\tau)\hat{m}_z(0) \rangle$ of Sr-doped NdNiO$_2$ as a functon of hole doping Sr $x$ calculated by DFT+DMFT at $T=290$ K.}
\label{fig:chi}
\end{figure}

Our DFT+DMFT calculations reveal a remarkable orbital-selective renormalization of the partially occupied Ni $x^2-y^2$ and $3z^2-r^2$ bands (shown in Fig.~\ref{fig:correlations}). In particular, for Sr $x=0$ the Ni $x^2-y^2$ states exhibit a large mass renormalization of $m$*/$m \sim 3$, while correlation effects in the $3z^2-r^2$ band are significantly weaker, $m$*/$m \sim 1.3$. 
This behavior is consistent with sufficiently different occupations of the Ni $x^2-y^2$ and $3z^2-r^2$ orbitals. In fact, the $x^2-y^2$ orbital occupancy for Sr $x=0$ is close to half-filling ($\sim$0.58 per spin-orbit), while the $3z^2-r^2$ orbitals are nearly fully occupied ($\sim$0.84). 
In addition, our analysis of the local spin susceptibility $\chi(\tau)=\langle \hat{m}_z(\tau)\hat{m}_z(0) \rangle$ (see Fig.~\ref{fig:chi}) suggests the proximity of the Ni $x^2-y^2$ states to localization, while the Ni $3z^2-r^2$ electrons are delocalized. Indeed, $\chi(\tau)$ for the Ni $3z^2-r^2$ states is seen to decay fast to zero with the imaginary time $\tau$, which is typical for itinerant behavior. In contrast to that $\chi(\tau)$ for the $x^2-y^2$ states is sufficiently larger, $\chi(0)=0.72$ $\mu_\mathrm{B}^2$, slowly decaying to $\sim$0.07 $\mu_\mathrm{B}^2$ at $\tau=\beta/2$.
Our results therefore suggest that magnetic correlations in NdNiO$_2$ are at the verge of orbital-dependent formation of local magnetic moments \cite{PhysRevB.101.041104}. In agreement with this the calculated (instantaneous) magnetic moment of Ni is about $\sqrt{\langle \hat{m}_z^2\rangle} \simeq 1.1$ $\mu_\mathrm{B}$, which is consistent with nearly a $S=1/2$ state of nickel. 

Upon hole doping the Ni $3d$ occupations slightly decrease to 0.52 and 0.80 (per spin-orbital) for the Ni $x^2-y^2$ and $3z^2-r^2$ orbitals, respectively, for Sr $x=0.5$. This corresponds to a $\sim$0.17 decrease of the total Wannier Ni $3d$ occupation, whereas the Nd $5d$ and O $2p$ states occupancies drop by $\sim$0.21 and 0.06.
In addition, we observe a gradual decrease of mass renormalization of the $x^2-y^2$ states to $m$*/$m \sim 2.3$ at Sr $x=0.5$. In contrast to that for the $3z^2-r^2$ orbital $m$*/$m$ slightly increases to $\sim$1.5. 
We notice no qualitative change in the self-energy upon changing of the Sr content $x$. The Ni $3d$ states obey a Fermi-liquid-like behavior with a weak damping at the Fermi energy.
Moreover, doping with Sr does not affect much magnetic moments in the paramagnetic phase of (Nd,Sr)NiO$_2$. Thus, the instantaneous magnetic moments $\sqrt{\langle \hat{m}^2_z \rangle}$ tend to increase only by about 5\%. Interestingly in our model calculations (with absent self-consistency over the charge density, i.e., for the fixed tight-binding parameters of the DFT Wannier Hamiltonian) this increase is more significant, about 63\%, suggesting the proximity to spin freezing, in accordance to recent model DMFT calculations \cite{PhysRevB.101.041104}. 

\begin{figure}[h]
\includegraphics[width=0.45\textwidth]{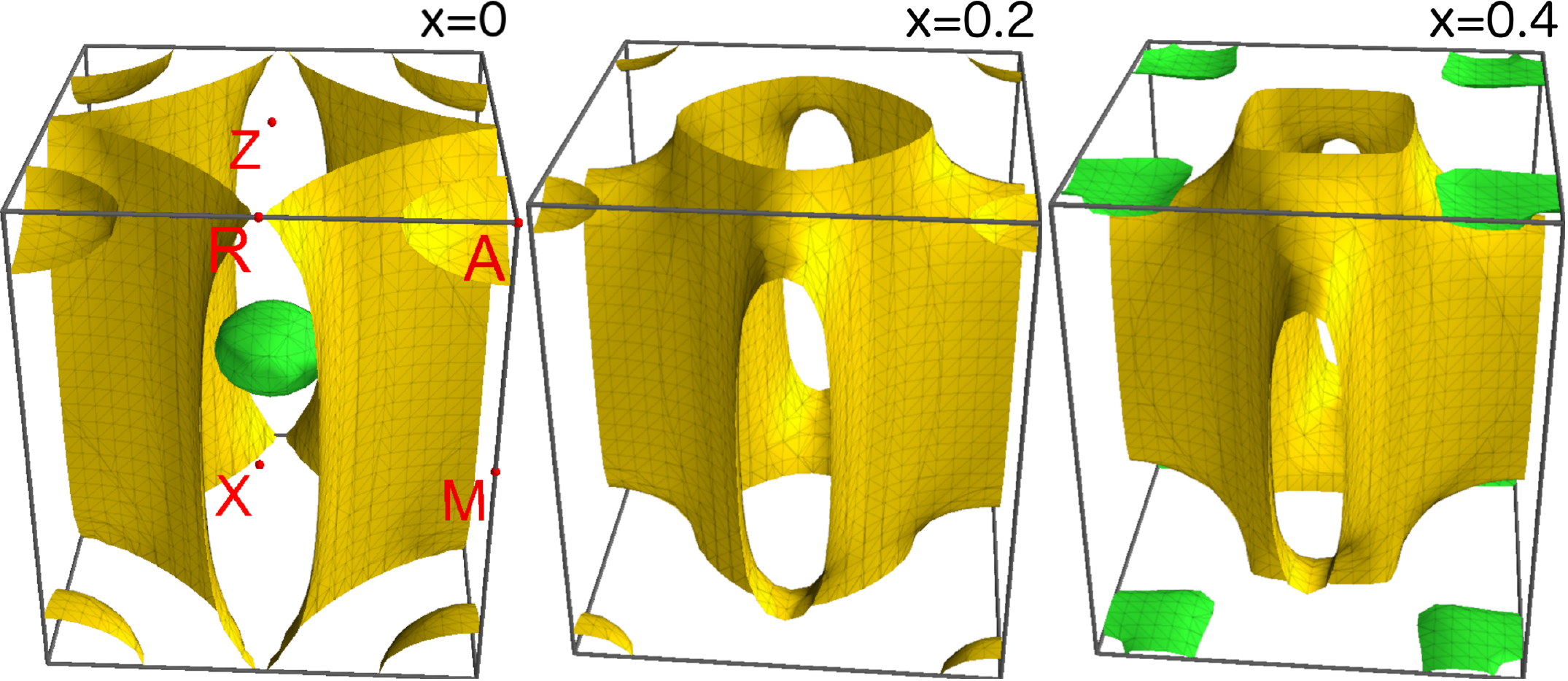}
\caption{Quasiparticle Fermi surface of Sr-doped NdNiO$_2$ for Sr $x=0$, 0.2, and 0.4 calculated by DFT+DMFT for the paramagnetc state at $T=290$ K.}
\label{fig:FS}
\end{figure}

Next, we calculate the quasiparticle Fermi surface (FS) of PM NdNiO$_2$ within DFT+DMFT. In Fig.~\ref{fig:FS} we show the dependence of the calculated FS's as a function of Sr $x$. We note that our results for $x=0$ are in qualitative agreement with previous band-structure studies \cite{Anisimov+Pickett,DFT_appl}. In particular, we obtain that the FS consists of three FS sheets, with the elliptical FS centered at the Brilloun zone (BZ) center ($\Gamma$ point), originating from the mixed Ni $3d$ and Nd $3z^2-r^2$ states. The electron FS pockets centered at the $A$-point are mainly of the Ni $xz/yz$ character. Similar to the cuprates, the FS of NdNiO$_2$ is dominated by the quasi-two-dimensional (quasi-2D) holelike FS sheet with a predominant Ni $x^2-y^2$ character, centered at the $A$-$M$ BZ edge.
In close similarity to the cuprates, our results for the FS topology imply an in-plane nesting with magnetic vector $q_m=(110)$ ($M$-point). 

Upon increase of the Sr content, we observe a remarkable change of the electronic structure of 
(Nd,Sr)NiO$_2$ which is associated with an entire reconstruction of the FS topology, i.e., a 
Lifshitz transition. Thus, at $x=0.2$ the elliptical FS centered at the $\Gamma$ point vanishes. 
In addition, the holelike quasi-2D FS sheets at the top and the bottom of the BZ merge near the $R$ 
point to a quasi-3D electron-like FS that forms a neck at the top and the bottom of the BZ. Overall, 
this suggests that the Lifshitz transition is accompanied by a reconstruction of magnetic correlations 
in infinite-layer (Nd,Sr)NiO$_2$ that appears near to the experimentally observed doping Sr $x\simeq 0.2$.

\begin{figure}[h]
\includegraphics[width=0.5\textwidth]{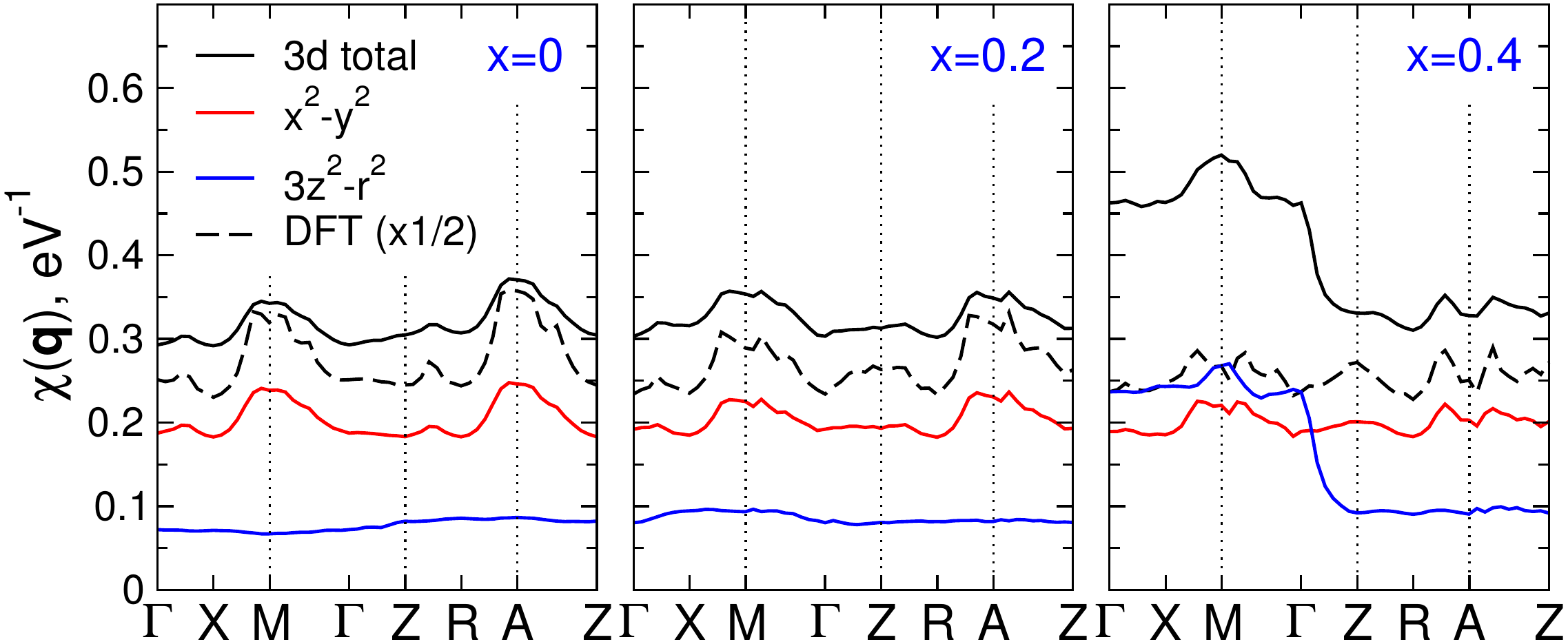}
\caption{Orbitally resolved static spin susceptibility $\chi({\bf q})$ of Sr-doped NdNiO$_2$ calculated 
by DFT+DMFT at $T=290$ K.}
\label{fig:chiq}
\end{figure}

\begin{figure}[h]
\includegraphics[width=0.4\textwidth]{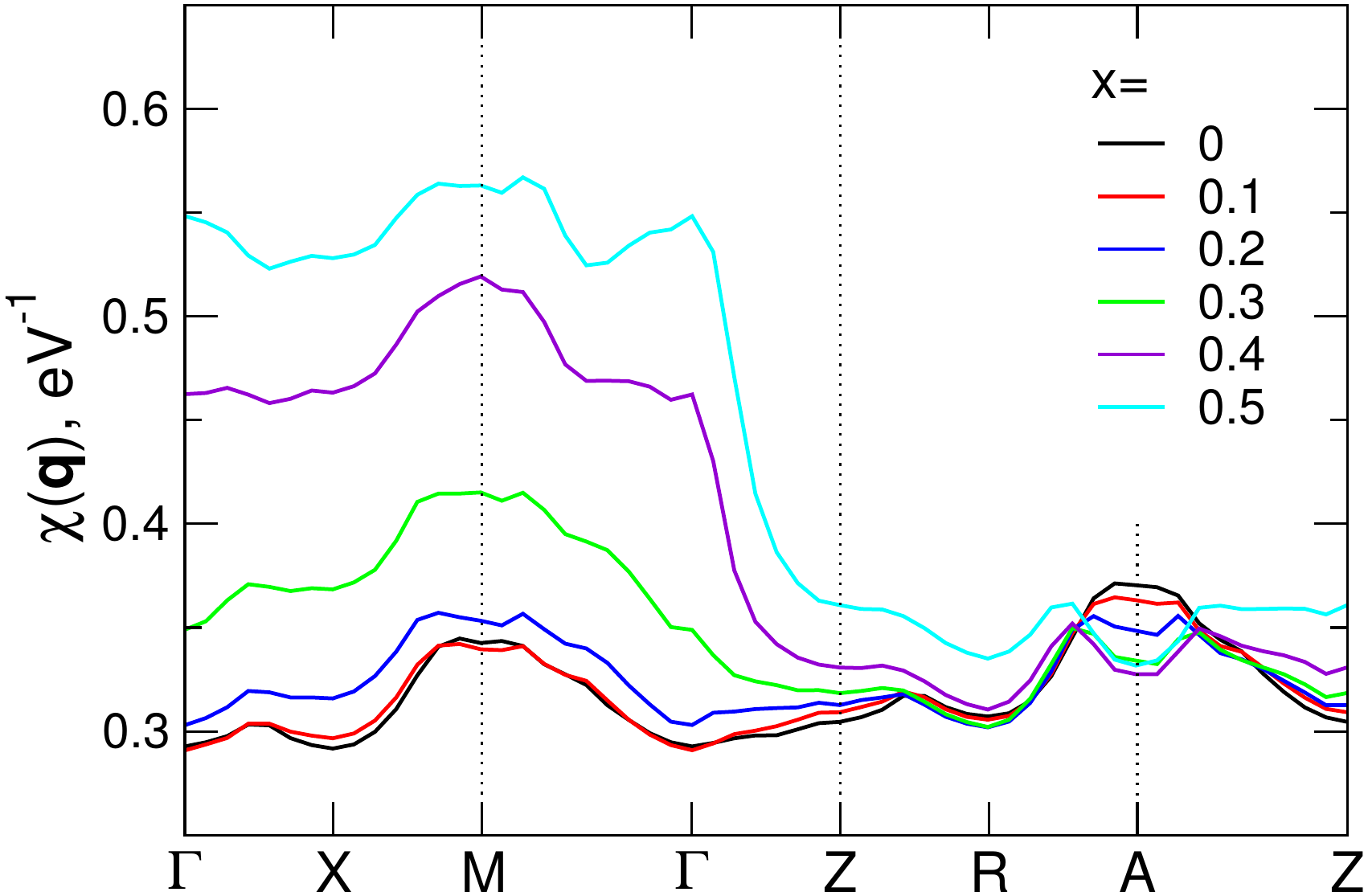}
\caption{Static spin susceptibility $\chi({\bf q})$ of Sr-doped NdNiO$_2$ as a functon of hole doping Sr $x$ calculated by DFT+DMFT at $T=290$ K.}
\label{fig:chiq_total}
\end{figure}

\begin{figure}[h]
\includegraphics[width=0.4\textwidth]{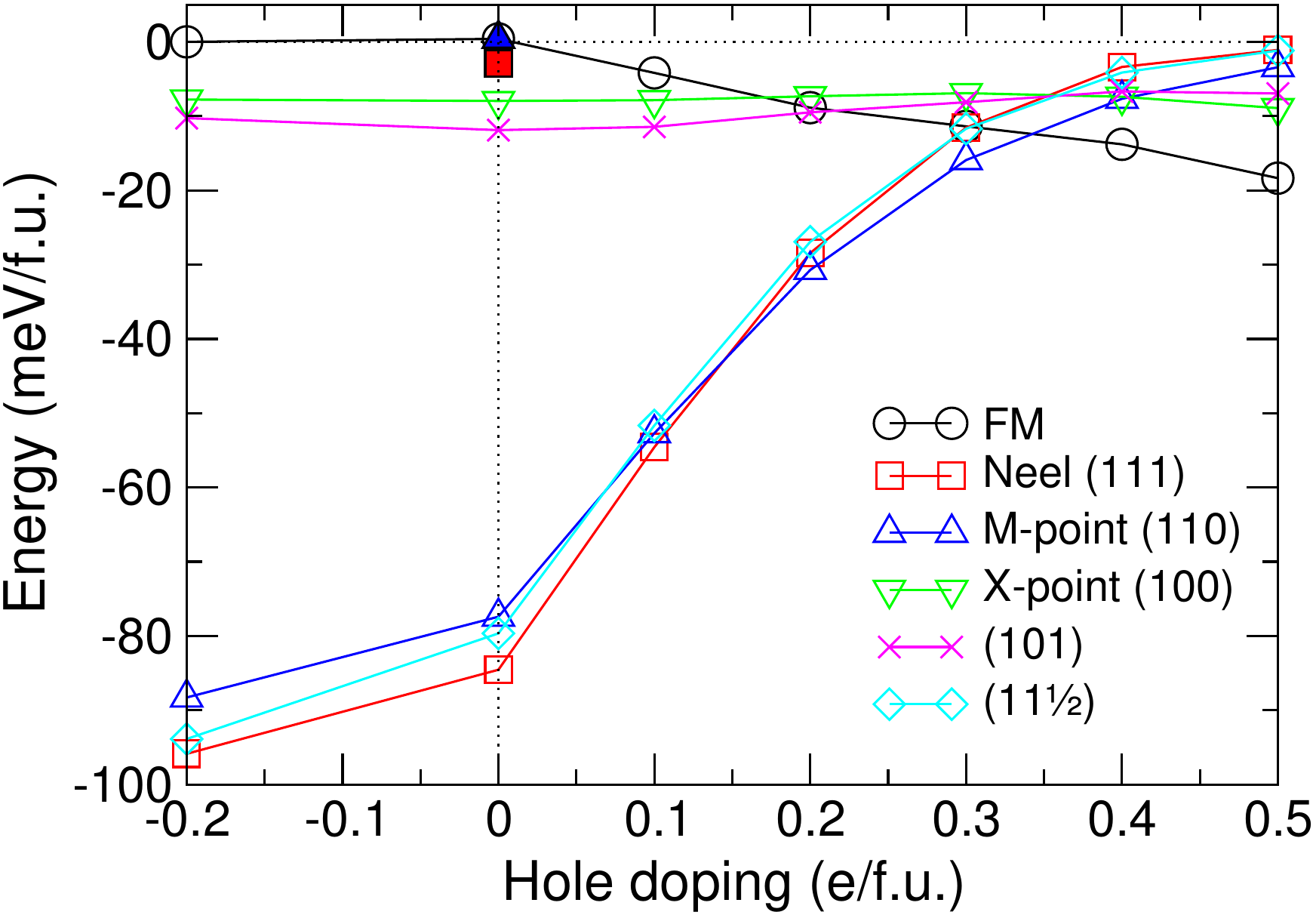}
\caption{Total energy difference $\Delta E = E_\mathrm{mag.}-E_\mathrm{NM}$ between the long-range 
magnetically ordered and non-magnetic states of NdNiO$_2$ as a function of hole doping calculated by 
DFT (empty symbols). Ferromagnetic (FM), N\'eel $(111)$, $C$-type $(110)$, $(101)$ and single stripe $(100)$, and 
staggered dimer $(11\frac{1}{2})$ states are shown. DFT+DMFT results for the total energy difference 
between the N\'eel and $C$-type AFM states and the paramagnetic state at $T=290$ K are depicted by 
filled symbols.}
\label{fig:etot}
\end{figure}

We proceed with analysis of the symmetry and strength of magnetic correlations in (Nd,Sr)NiO$_2$. 
For this purpose we compute the momentum-dependent static magnetic susceptibility $\chi({\bf q})$ 
within DFT+DMFT using the particle-hole bubble approximation. Orbital contributions of $\chi({\bf q})$ 
along the BZ path and their dependence on the Sr content $x$ are shown in Fig.~\ref{fig:chiq}. 
Our results for the total $\chi({\bf q})$ as a function of Sr doping $x$ are summarized in 
Fig.~\ref{fig:chiq_total}. Interestingly for $x=0$ our results for $\chi({\bf q})$ exhibit two well 
defined maxima at the $M$ and $A$ points of the tetragonal BZ. This suggests the existence of 
(at least) two leading magnetic instabilities due to the Ni $x^2-y^2$ states (for $x=0$) with a wave 
vector near to $q_m=(110)$ and $(111)$, that corresponds to the $C$-type and the N\'eel AFM ordering, 
respectively. In the same time $\chi({\bf q})$ for the $3z^2-r^2$ states is seen to be small and nearly 
{\bf q}-independent. We notice that $\chi({\bf q})$ appears to be somewhat 
larger in the $A$ than that in the $M$ point. We therefore expect that the three-dimensional N\'eel 
AFM state is more energetically favorable than the quasi-2D $C$-type (for Sr $x=0$). 
In fact, this qualitative analysis agrees well with our total-energy calculations within the 
spin-polarized DFT and DFT+DMFT methods (see Fig.~\ref{fig:etot}). Both reveal that for Sr $x=0$ 
the N\'eel AFM ordering is more energetically favorable by about 3-4 meV/f.u. with respect to the 
$C$-type AFM and the PM state within DFT+DMFT, at $T=290$~K. We note that within DFT 
the N\'eel and the staggered dimer $(11\frac{1}{2})$ 
and $C$-type (110) states are differ by about 5-7 meV/f.u., while the non-magnetic state appears much above, by about 85 meV/f.u.

Our results for $\chi({\bf q})$ and total energies suggest that various types 
of spin order are competing (nearly energetically degenerate) in (Nd,Sr)NiO$_2$. Indeed, for Sr $x=0.2$, $\chi({\bf q})$
is seen to be nearly flat and degenerate at around the $M$ and $A$ points (see Fig.~\ref{fig:chiq}), 
implying possible frustration of the Ni $3d$ moments. Upon further increase of Sr $x$, our results 
provide a clear evidence of an entire reconstruction of magnetic correlations, with the $3z^2-r^2$ states 
now playing a major role. While for Sr $x=0.4$ $\chi({\bf q})$ for the $x^2-y^2$ orbital is seen to be 
nearly flat (degenerate for different {\bf q}), suggesting in-plane frustration of the Ni $3d$ moments. 
The out-of-plane $3z^2-r^2$ orbital contribution reveals a flat maximum near the $M$ point.

\begin{figure}[h]
\includegraphics[width=0.4\textwidth]{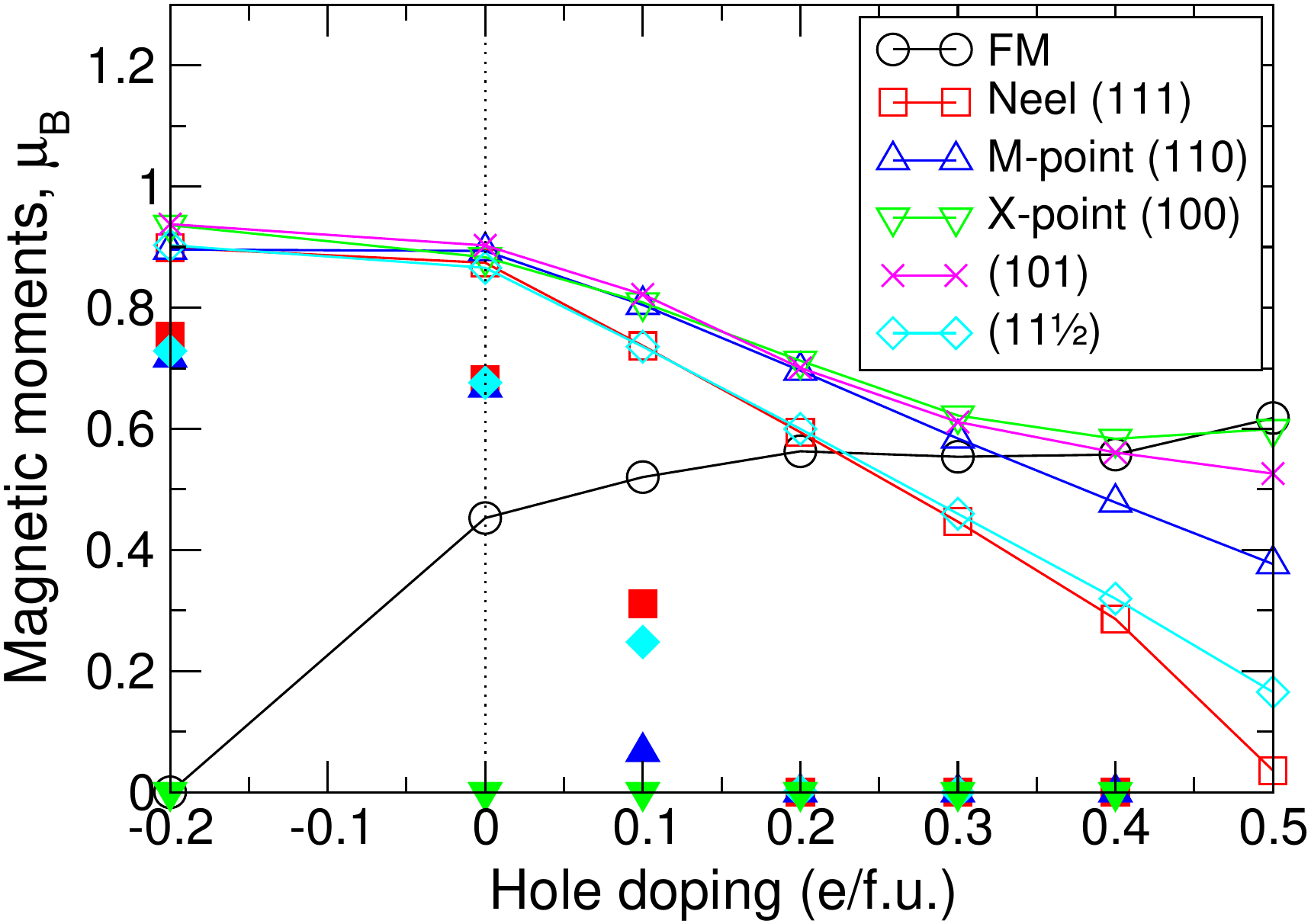}
\caption{Long-range ordered magnetic moments of Ni as a function of hole doping calculated for NdNiO$_2$ by DFT (empty symbols). DFT+DMFT results for the N\'eel, $C$-type $(110)$, single stripe $(100)$, and staggered dimer $(11\frac{1}{2})$ AFM states at $T=290$ K are shown by filled symbols.}
\label{fig:mag}
\end{figure}

In Fig.~\ref{fig:mag} we show our results for the long-range ordered magnetic moments of nickel 
calculated within the spin-polarized DFT and DFT+DMFT. The latter are about 0.67~$\mu_\mathrm{B}$/Ni as 
obtained by DFT+DMFT for the N\'eel $(111)$, $C$-type $(110)$, and staggered dimer $(11\frac{1}{2})$ 
AFM states for Sr $x=0$, at $T=290$ K. Notably, we observe a sharp suppression of the calculated 
magnetization $M_z$ and hence of the N\'eel temperature evaluated from the spin-polarized DFT+DMFT 
calculations with Sr $x$. In particular, for Sr $x=0.2$ we find no evidence for a magnetically ordered state at 
$T \geq 290$~K. Thus, all magnetic configurations discussed here, namely, the $(100)$, $(110)$, $(111)$ 
and $(11\frac{1}{2})$ AFM and FM configurations collapse in the PM state. That is, for Sr $x=0.2$ the 
N\'eel (Curie) temperature is much below the room temperature that suggests rising of quantum spin 
fluctuations with $x$.

We note, however, that analysis of the finite-temperature DFT+DMFT results may often be problematic. 
For example, the single stripe (100) AFM and ferromagnetic orderings are found to be unstable at $T=290$~K, 
i.e, both collapse to the PM state. We therefore first perform the spin-polarized DFT calculations of the 
ground state energy differences between different magnetic states (see Fig.~\ref{fig:etot}). 
In fact, the DFT calculations give qualitatively similar results to those obtained by DFT+DMFT with significantly larger values of the total energy difference (with respect to the non-magnetic state) of $\sim$85 meV/f.u., and 8 meV/f.u. for the $C$-type, N\'eel, and staggered dimer $(11\frac{1}{2})$, and single stripe $(100)$ magnetic states, respectively. 
Both spin-polarized DFT and DFT+DMFT calculations reveal a near degeneracy of various types of spin orders, implying frustration of magnetic correlations in (Nd,Sr)NiO$_2$. The latter is most notable for the Sr content of about $x \simeq 0.2$--0.3, which is close to the experimental Sr doping $x \simeq0.2$. Moreover, the calculated magnetization for the various AFM states tends to decrease in both the DFT and DFT+DMFT calculations upon increase of Sr $x$ (see Fig.~\ref{fig:mag}). In addition, within DFT+DMFT magnetization is found to sharply collapse to the PM state for Sr $x>0.1$ at $T=290$~K, suggesting a sharp increase of spin fluctuations with $x$.

\begin{figure}[h]
\includegraphics[width=0.4\textwidth]{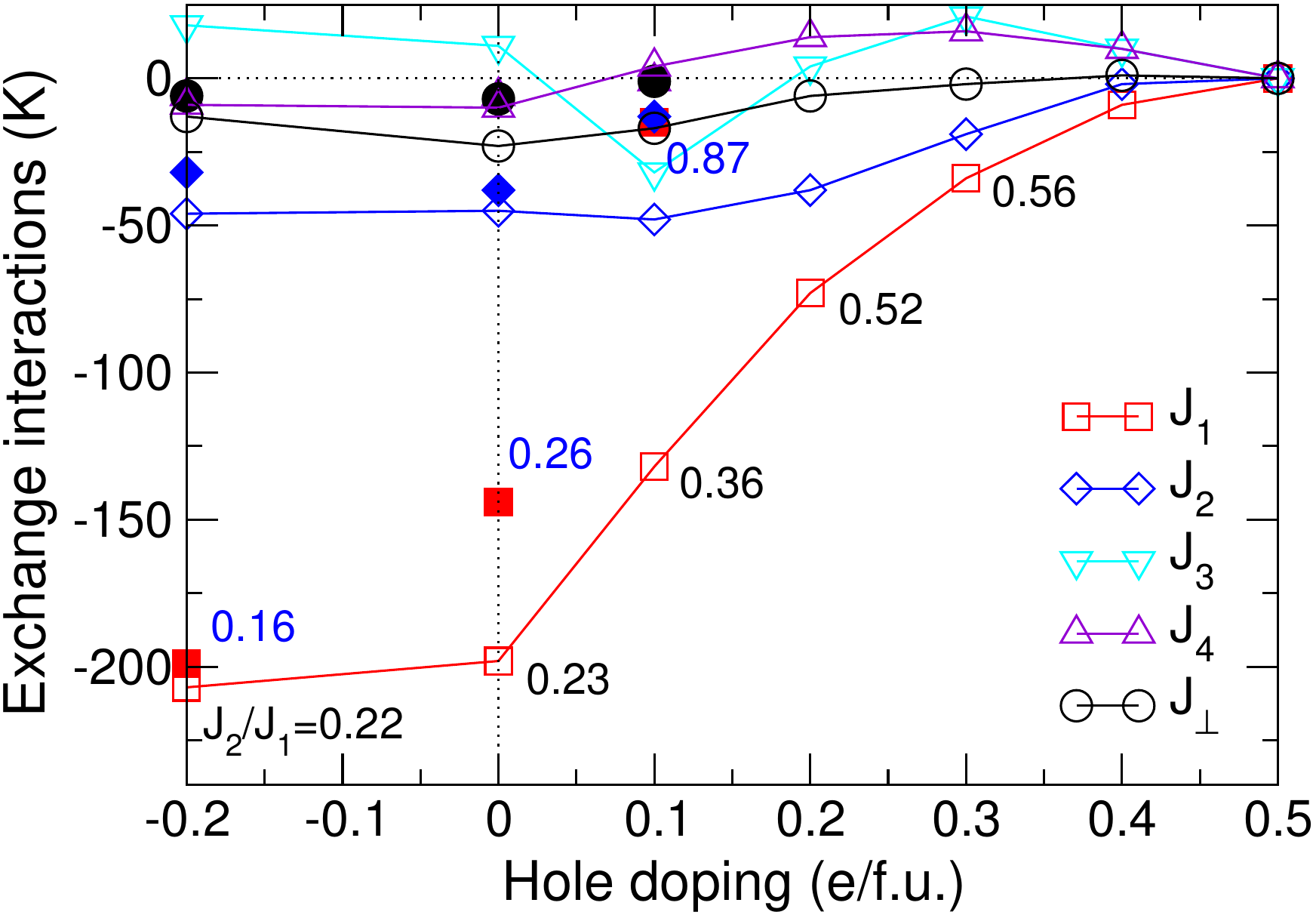}
\caption{Exchange interaction parameters (in-plane nearest-neighbor (NN) $J_1$, next-nearest neighbor $J_2$, 3-rd NN $J_3$, 4-th NN $J_4$ and interlayer coupling $J_{\perp}$) of Sr-doped NdNiO$_2$ calculated within spin-polarized DFT (empty symbols) and DFT+DMFT (filled symbols).}
\label{fig:exch}
\end{figure}

Our results point out an anomalous sensitivity of the electronic structure and magnetic correlations of 
(Nd,Sr)NiO$_2$ with respect to the Sr $x$ doping. In particular, we found a remarkable frustration of 
(orbital-dependent) magnetic moments of Ni sites near to the optimal Sr doping $x \simeq0.2$. To help 
check these results, we computed magnetic exchange couplings within the spin-polarized DFT and DFT+DMFT using the magnetic force theorem \cite{exchange}. 
Our findings for the N\'eel AFM state are summarized in Fig.~\ref{fig:exch}. We observe that for Sr $x=0$ the interlayer couplings $J_\perp$ is small and weakly antiferromagnetic, $J_\perp \sim -23$ K \cite{Heisenberg_coeff}. The in-plane couplings $J_1$ (nearest-neighbor) and $J_2$ 
(next-nearest-neighbor) are both antiferromagnetic and are sufficiently higher by modulus, $\sim$-198 K and -45 K, respectively. Interestingly that for pure NdNiO$_2$, 
$|J_1| = 198$ K is comparable to that found experimentally in infinite-layer CaCuO$_2$ \cite{NatPhys.13.1201}. Most importantly, our results reveal a remarkable change of the $J_2/J_1$ ratio 
with respect to Sr $x$, which is increasing from $\sim$0.36 to 0.56 for $x=0.1$--0.3, i.e., near to the 
experimental doping Sr $x \simeq 0.2$. While in DFT+DMFT magnetization is found to quickly collapse to the PM state for Sr $x>0.1$, the exchange couplings evaluated from the spin-polarized DFT+DMFT calculations do follow the same trend, with $J_2/J_1 \simeq 0.26$ for Sr $x=0$, which is found to increase to 0.87 for Sr $x=0.1$.

Our findings resemble us the behavior of the spin-1/2 frustrated $J_1$-$J_2$ Heisenberg model on the 
two-dimensional (2D) square lattice, with an unusual quantum spin liquid ground state to appear in the 
highly frustrated region $J_2/J_1 \simeq 0.4$--0.5, sandwiched 
between the N\'eel and stripe type (or valence-bond solid) ordered states \cite{Heisenberg_model}.
This analogy is very striking, taking into account our results for the change of the electronic structure 
and magnetic couplings $J_2/J_1$ ratio in (Nd,Sr)NiO$_2$ with Sr $x$. Thus, the frustration region is sandwiched between 
the two different (long- or short-range ordered) antiferromagnets \cite{Heisenberg_model}.
We find that magnetic couplings in (Nd,Sr)NiO$_2$ near to the optimal doping demonstrate an unanticipated 
frustration, which suppresses a long-range magnetic order (resulting in a drastic drop of the N\'eel temperature), and can lead to formation of unusual quantum 
spin liquid ground states.
Moreover, our results suggest that frustration is maximal for Sr-doping $x=0.1$--0.2 that corresponds to the highly frustrated region of the spin-1/2 frustrated $J_1$-$J_2$ Heisenberg model. 
Overall, our results suggest the importance of in-plane spin fluctuations to explain superconductivity in (Nd,Sr)NiO$_2$, in contrast to the previous claims \cite{Nature.572.624}.
We point out that strong frustration of magnetic interactions in (Nd,Sr)NiO$_2$ suggests that superconductivity in infinite-layer (Nd,Sr)NiO$_2$ appears to be similar to that observed in iron chalcogenides and pnictides \cite{FSCs_frustration}.

In conclusion, we employed the DFT+DMFT computational approach to study the effects of 
electronic correlations and Sr-doping on the electronic structure and magnetic properties of 
(Nd,Sr)NiO$_2$.
We show that upon hole doping it undergoes a Lifshitz transition 
of the Fermi surface which is accompanied by a reconstruction of magnetic correlations.
Most importantly, magnetic interactions in (Nd,Sr)NiO$_2$ are found to demonstrate an unanticipated frustration.
We find that frustration is maximal for Sr-doping $x=0.1$--0.2 that nearly corresponds to the experimentally observed doping value of (Nd,Sr)NiO$_2$. 
Our results for (Nd,Sr)NiO$_2$ reveal a feature that is central to copper oxides as well as to iron chalcogenides and pnictides -- large in-plane spin fluctuations. We propose that superconductivity in nickelates is strongly influenced, or even induced, by in-plane spin fluctuations.

\begin{acknowledgments}

We acknowledge support by the state assignment of Minobrnauki of Russia (theme ``Electron'' No. AAAA-A18-118020190098-5). Theoretical analysis of the electronic structure and Fermi surface topology was supported by Russian Foundation for Basic Research (Project No. 18-32-20076). S.Y.S. was supported by National Science Foundation DMR Grant No. 1832728.

\end{acknowledgments}

\pagebreak

\section*{Supplementary Material}
\beginsupplement

\begin{thebibliography}{100}

\bibitem{Nature.572.624}
D. Li, K. Lee, B. Y. Wang, M. Osada, S. Crossley \emph{et al.}, Nature (London) \textbf{572}, 624 (2019).

\bibitem{Nature.356.775}
M. Azuma, Z. Hiroi, M. Takano, Y. Bando, and Y. Takeda, Nature (London) 356, 775 (1992).

\bibitem{NatPhys.13.1201}
Y. Y. Peng, G. Dellea, M. Minola, M. Conni, A. Amorese \emph{et al.}, Nat. Phys. \textbf{13}, 1201 (2017).

\bibitem{PhysRevLett.77.4430} S. Y. Savrasov and O. K. Andersen, Phys. Rev. Lett. \textbf{77}, 4430 (1996).


\bibitem{Anisimov+Pickett}
V. I. Anisimov, D. Bukhvalov, T. M. Rice, Phys. Rev. B \textbf{59}, 7901 (1999);
K.-W. Lee and W. E. Pickett, Phys. Rev. B \textbf{70}, 165109 (2004);
M.-Y. Choi, K.-W. Lee, and W. E. Pickett, Phys. Rev. B \textbf{101}, 020503(R) (2020).

\bibitem{NatMater2020}
M. Hepting \emph{et al.}, Nat. Mater. https://doi.org/10.1038/s41563-019-0585-z (2020).





\bibitem{DFT_appl}
P. Jiang, L. Si, Z. Liao, and Z. Zhong, Phys. Rev. B \textbf{100}, 201106 (2019);
Y. Nomura, M. Hirayama, T. Tadano, Y. Yoshimoto, K. Nakamura, and R. Arita, Phys. Rev. B \textbf{100}, 205138 (2019);
M. Jiang, M. Berciu, and G. A. Sawatzky, arXiv: 1909.02557 (2019);
J. Gao, Z. Wang, C. Fang, and H. Weng, arXiv: 1909.04657 (2019);
H. Zhang, L. Jin, S. Wang, B. Xi, X. Shi, F. Ye, and J.-W. Mei, arXiv: 1909.07427 (2019);
E. Been, W.-S. Lee, H. Y. Hwang, Y. Cui, J. Zaanen \emph{et al.}, arXiv:2002.12300 (2020).

\bibitem{PhysRevX.10.011024}
A. S. Botana and M. R. Norman, Phys. Rev. X. \textbf{10}, 011024 (2020);

\bibitem{Geisler_arxiv_2020}
B. Geisler and R. Pentcheva, arXiv: 2001.03762

\bibitem{Models_appl}
G.-M. Zhang, Y.-F. Yang, and F.-C. Zhang, Phys. Rev. B \textbf{101}, 020501(R) (2020);
X. Wu, D. Di Sante, T. Schwemmer, W. Hanke, H. Y. Hwang, S. Raghu, and R. Thomale, arXiv: 1909.03015 (2019).

\bibitem{Sakakibara_arxiv_2019}
H. Sakakibara, H. Usui, K. Suzuki, T. Kotani, H. Aoki, and K. Kuroki, arXiv: 1909.00060 (2019).

\bibitem{dmft}
A. Georges, G. Kotliar, W. Krauth, and M. J. Rozenberg, Rev. Mod. Phys. \textbf{68}, 13 (1996);
G. Kotliar, S. Y. Savrasov, K. Haule, V. S. Oudovenko, O. Parcollet, and C. A. Marianetti, Rev. Mod. Phys. \textbf{78}, 865 (2006).


\bibitem{dft+dmft}
K. Haule, Phys. Rev. B \textbf{75}, 155113 (2007);
L. V. Pourovskii, B. Amadon, S. Biermann, and A. Georges,
Phys. Rev. B \textbf{76}, 235101 (2007);
B. Amadon, F. Lechermann, A. Georges, F. Jollet, T. O. Wehling, and A. I. Lichtenstein,  Phys. Rev. B \textbf{77}, 205112 (2008);
M. Aichhorn, L. Pourovskii, V. Vildosola, M. Ferrero, O. Parcollet \emph{et al.}, Phys. Rev. B  \textbf{80}, 085101 (2009);
I. Leonov, L. Pourovskii, A. Georges, and I. A. Abrikosov,
Phys. Rev. B \textbf{94}, 155135 (2016).


\bibitem{PhysRevB.101.041104}
Ph. Werner and S. Hoshino, Phys. Rev. B \textbf{101}, 041104(R) (2020).

\bibitem{Ryee_arxiv_2019}
S. Ryee, H. Yoon, T. J. Kim, M. Y. Jeong, and M. J. Han, arXiv: 1909.05824 (2019).

\bibitem{Gu_arxiv_2019}
Y. Gu, S. Zhu, X. Wang, J. Hu, and H. Chen, arXiv: 1911.00814 (2019).

\bibitem{Held_group}
L. Si, W. Xiao, J. Kaufmann, J. M. Tomczak, Y. Lu, Z. Zhong, and K. Held, arXiv: 1911.06917 (2019);
M. Kitatani, L. Si, O. Janson, R. Arita, Z. Zhong, K. Held, arXiv: 2002.12230 (2020).

\bibitem{Lechermann_arxiv_2019}
F. Lechermann, arXiv: 1911.11521 (2019).

\bibitem{Karp_arxiv_2020}
J. Karp, A. S. Botana, M. R. Norman, H. Park, M. Zingl, A. Millis, arXiv: 2001.06441 (2020).


\bibitem{pseudo}
P. Giannozzi, S. Baroni, N. Bonini, M. Calandra, R. Car \emph{et al.}, J. Phys.: Condens. Matter \textbf{21}, 395502 (2009).

\bibitem{KCuF3}
I. Leonov, N. Binggeli, Dm. Korotin, V. I. Anisimov, N. Stoji\'c, and D. Vollhardt, Phys. Rev. Lett. \textbf{101}, 096405 (2008);
I. Leonov, Dm. Korotin, N. Binggeli, V. I. Anisimov, and D. Vollhardt,
Phys. Rev. B \textbf{81}, 075109 (2010).

\bibitem{dftdmft_application}
Z. P. Yin, K. Haule, and G. Kotliar, Nat. Mater. {\bf 10}, 932 (2011); Nat. Phys. {\bf 7}, 294 (2011);
%
L. de' Medici, J. Mravlje, and A. Georges, Phys. Rev. Lett. {\bf 107}, 256401 (2011);
%
P. Werner, M. Casula, T. Miyake, F. Aryasetiawan, A. J. Millis, and S. Biermann, 
Nat. Phys. {\bf 8}, 331 (2012);
%
I. Leonov, S. L. Skornyakov, V. I. Anisimov, and D. Vollhardt,
Phys. Rev. Lett. {\bf 115}, 106402 (2015);
%
S. L. Skornyakov, V. I. Anisimov, D. Vollhardt, and I. Leonov,
Phys. Rev. B {\bf 96} 035137 (2017);
%
P. V. Arribi and L. de' Medici,
Phys. Rev. Lett. {\bf 121}, 197001 (2018);
%
E. Greenberg, I. Leonov, S. Layek, Z. Konopkova, M. P. Pasternak \emph{et al.}, Phys. Rev. X {\bf 8}, 031059 (2018);
%
I. Leonov, G.K. Rozenberg, I.A. Abrikosov, npj Comput. Mater. {\bf 5}, 90 (2019);
%
X. Deng, K. M. Stadler, K. Haule,  A. Weichselbaum, J. von Delft, and G. Kotliar, 
Nat. Commun. {\bf 10}, 2721 (2019).

\bibitem{Wannier}
N. Marzari, A. A. Mostofi, J. R. Yates, I. Souza, and D. Vanderbilt, Rev. Mod. Phys. \textbf{84}, 1419 (2012);
%
V. I. Anisimov, D. E. Kondakov, A. V. Kozhevnikov, I. A. Nekrasov, Z. V. Pchelkina \emph{et al.}, Phys. Rev. B \textbf{71}, 125119 (2005).

\bibitem{CT-QMC}
E. Gull, A. J. Millis, A. I. Lichtenstein, A. N. Rubtsov, M. Troyer, and P. Werner, Rev. Mod. Phys. \textbf{83}, 349 (2011).

\bibitem{Park2014}
H. Park, A. J. Millis, and C. A. Marianetti, Phys. Rev. B \textbf{89}, 245133 (2014);
E. A. Nowadnick, J. P. Ruf, H. Park, P. D. C. King, D. G. Schlom \emph{et al.}, Phys. Rev. B \textbf{92}, 245109 (2015);
I. Leonov, A. S. Belozerov, and S. L. Skornyakov, Phys. Rev. B \textbf{100}, 161112(R) (2019)

\bibitem{exchange}
A. I. Liechtenstein, M. I. Katsnelson, V. P. Antropov, and V. A. Gubanov, 
J. Magn. Magn. Mater. {\bf 67}, 65 (1987);
Y. O. Kvashnin, O. Gr\r{a}n\"as, I. Di Marco, M. I. Katsnelson, A. I. Lichtenstein, and O. Eriksson, Phys. Rev. B \textbf{91}, 125133 (2015).

\bibitem{Heisenberg_coeff}
Here we adopt the following notation for the Heisenberg model $H=-\sum_{ij}J_{ij}e_ie_j$ where $e_{i,j}$ are the unit vectors.


\bibitem{Heisenberg_model}
S.-S. Gong, W. Zhu, D. N. Sheng, O. I. Motrunich, and M. P. A. Fisher,
Phys. Rev. Lett. \textbf{113}, 027201 (2014);
S. Morita, R. Kaneko, and M. Imada, J. Phys. Soc. Jpn. {\bf 84}, 024720 (2015);
L. Wang and A. W. Sandvik, Phys. Rev. Lett. \textbf{121}, 107202 (2018).

\bibitem{FSCs_frustration}
Q. Si and E. Abrahams, Phys. Rev. Lett. \textbf{101}, 076401 (2008);
C. Fang, H. Yao, W.-F. Tsai, J. P. Hu, and S. A. Kivelson, Phys. Rev. B \textbf{77}, 224509 (2008);
C. Xu, M. M\"uller, and S. Sachdev, Phys. Rev. B 78, 020501(R) (2008);
M. J. Han, Q. Yin, W. E. Pickett, and S. Y. Savrasov, Phys. Rev. Lett. \textbf{102}, 107003 (2009);
J. K. Glasbrenner, I. I. Mazin, H. O. Jeschke, P. J. Hirschfeld, R. M. Fernandes, and R. Valent\'i, Nat. Phys. \textbf{11}, 953 (2015);
A. Baum, H. N. Ruiz, N. Lazarevi\'c, Y. Wang, T. B\"ohm \emph{et al.},
Commun. Phys. \textbf{2}, 14 (2019).




\end{thebibliography}
\end{document}